\documentclass[]{aastex62}
\usepackage{amssymb,amsmath}

\def\eq#1{\begin{equation} #1 \end{equation}}

\def\atm    {\hbox{\rm{ATM}}}

\graphicspath{{./}{figures/}}

\received{July 13, 2020}
\accepted{--}
\submitjournal{Icarus}

\shorttitle{Asteroid Size Estimation with LSST data}
\shortauthors{Ivezi\'{c} \& Ivezi\'{c}}

\begin{document}

\title{Predicting the Accuracy of Asteroid Size Estimation with Data from the Rubin Observatory Legacy Survey of Space and Time}

\correspondingauthor{\v{Z}eljko Ivezi\'{c}}
\email{ivezic@uw.edu}

\author[0000-0001-5986-8929]{Vedrana Ivezi\'{c}}
\affiliation{Department of Computer Science, Princeton University, 35 Olden St, Princeton, NJ 08540, USA} 

\author[0000-0001-5250-2633]{\v{Z}eljko Ivezi\'{c}}
\affiliation{Department of Astronomy and the DiRAC Institute, University of Washington, 3910 15th Avenue NE, Seattle, WA 98195, USA}

\begin{abstract}
Recent work  has shown that the correlation between SDSS colors and optical albedo can be used 
to estimate asteroid sizes from optical data alone.
We revisit a correlation between SDSS colors and optical albedo for asteroids, with the albedo
derived using WISE-based size estimates. Moeyens, Myhrvold \& Ivezi\'{c} (2020) showed 
that this correlation can be used to estimate asteroid sizes with optical data alone, with a precision 
of about 17\% relative to WISE-based size estimates. We present here several more 
sophisticated data-driven models for the variation of optical albedo with colors and estimate 
the contribution of SDSS photometric errors to the albedo and size estimate uncertainties. 
We use the results of our analysis to predict that LSST data will enable asteroid size precision 
of about 15\% relative to WISE-based size estimates. Compared to the accuracy of WISE-based 
size estimates of 15-20\%, the implied accuracy of optical size estimates, in the range 21-25\%,
is thus only a factor of 1.3 to 1.4 worse.  This size estimation accuracy is significantly better 
than commonly assumed for optical data and is due to accurate and homogeneous multi-band 
photometry delivered by modern digital sky surveys.  
\end{abstract}


\section{Introduction}

Methods for estimating asteroid sizes are important both in the context of studying the formation
and evolution of the asteroid belt as encoded in its size distribution \citep[e.g.,][]{MOC2008}, and 
in the context of planetary defense where the potential damage caused by an impactor scales with
its size \citep[e.g.,][]{NAP25476}. Despite this importance of asteroid size, there are fewer than 
a thousand asteroids with direct radar, occultation, and spacecraft size measurements 
(e.g., see figure 7 in \citealt{NM2018b}).

The largest asteroid sample with size estimates is based on thermal flux modeling and infrared fluxes 
measured by the WISE survey \citep{WISE}. A series of papers that produced size estimates for about 
164,000 asteroids, as well as constraints on asteroid emissivity properties, was reviewed and 
summarized by \cite{MainzerReview}. Recently, \cite{MMI}, hereafter MMI, published an open-source 
package for asteroid thermal flux modeling\footnote{See \url{https://github.com/moeyensj/atm}}, 
\atm, and used it to reproduce size estimates from the NEOWISE paper series with a scatter of only 
6\% and negligible bias (this scatter should not be confused with a precision or accuracy of 
size estimation because the same data were used in both studies -- rather, it represents the 
repeatability of results based on different modeling tools and methods; for more details, please see MMI). 
The internal precision (i.e., random, or statistical, errors) of WISE-based size estimates for the 
majority of sample is about 10\%, 
and 2-3\% for the brightest and best observed objects. The accuracy of WISE-based sizes, which 
combines precision and systematic errors, is estimated to be in the range 15-20\%. This estimate 
does not include unknown contributions of possibly over-simplified modeling framework, such as spherical 
asteroid approximation (\citealt{2016PDSS..247.....M}; see also MMI). We note that the behavior of 
systematic and random (statistical) uncertainties for the best-fit parameters remains an active research
topic \cite[see for example][]{2007astro.ph..3085W, 2018AJ....155...74M, 2018AJ.156.62M}. 

As demonstrated by MMI, asteroid size can also be estimated fairly precisely using the strong 
correlation between asteroid optical colors and optical albedo \citep{2012ApJ...745....7M}. 
Traditionally, optical size estimates were inferior to infrared-based size estimates 
because asteroid surfaces are not very shiny: their reflectivity (albedo) is low, which 
implies high emissivity via Kirchhoff's law \citep[for a detailed discussion, see][]{NM2018a}.
As a consequence, {\it dynamic range for optical albedo is much larger than dynamic range for
infrared emissivity}. 
For example, an uncertainty range in reflectivity of a factor of 2 around a fiducial value of 
0.1 corresponds to an emissivity uncertainty range around 0.9 of only $\sim$10\%. As a consequence, 
the corresponding size uncertainty range would be 41\% for optical estimates and $<5\%$ for 
infrared estimates, about an order of magnitude difference when other systematic error 
contributions are not taken into account. It turns out that without color information
the observed asteroid albedo distribution would result in a scatter of optical size 
estimates of about 50-60\%, and demonstrably non-Gaussian distribution (that is, 
a factor of 3 to 4 worse than the accuracy of infrared-based sizes). MMI showed that size 
estimates based on SDSS data can be tied to WISE-based estimates with an uncertainty of 17\%.
After adding in quadrature the 15-20\% accuracy estimated for WISE-based estimates (MMI),
the implied accuracy of optical size estimates is in the range 23-26\%, or only 
a factor of about 1.3 worse. 

This accuracy level for optical size estimates bodes well for future optical asteroid surveys,
such as the Rubin Observatory Legacy Survey of Space and Time \citep{2019ApJ...873..111I}, which 
might deliver such size estimates for over 5 million asteroids \citep[][and references 
therein]{2018Icar..303..181J}. MMI pointed out that their simplistic method
for mapping optical colors to albedo could be improved using modern machine learning methods.
Given the huge potential of LSST sample for asteroid studies, we revisit their result
here and present several more sophisticated data-driven models for the variation of optical
albedo with colors. In addition, we estimate the contribution of SDSS photometric errors 
to the albedo and size estimate uncertainties, and then scale the result to LSST data. 

The mapping of multi-dimensional color space to a scalar quantity, such as albedo, is not 
new to astronomy. There are numerous methods developed to handle mathematically analogous 
problems of estimating photometric redshifts for galaxies \citep[e.g.,][]{2018AJ....155....1G}
and photometric metallicity for stars \citep{2008ApJ...684..287I}. While physical reasons
for observed correlations with colors vary greatly, essentially the same analysis methods
can be brought to bear to the mapping of asteroid albedo with colors. 
After introducing the SDSS-WISE asteroid dataset, we discuss and apply several such methods 
in \S2. We summarize and discuss our results in \S3.   

All data files used in this work including WISE
and NEOWISE, Minor Planet Center, and SDSS Moving Object Catalog data are available for download
from this 
GitHub\footnote{See \url{https://github.com/ivezicV/2share/tree/master/AsteroidPaper}} site.
Motivated by 
a desire to enable transparency and reproducibility of results, we also publicly release Jupyter Notebook 
and supporting Python code used to perform analysis and produce all figures presented here.

\section{Optical Color-based Asteroid Size Estimation Methods \label{sec:methods}}

In this Section, we start by introducing the SDSS-WISE dataset used in our analysis and then 
describe the constant-albedo models, followed by the Nearest Neighbor method and 
Gaussian mixture models. 

\subsection{Description of SDSS-WISE dataset \label{sec:dataset}}

\begin{figure}[th]
\centering
\includegraphics[width=0.8\textwidth, keepaspectratio]{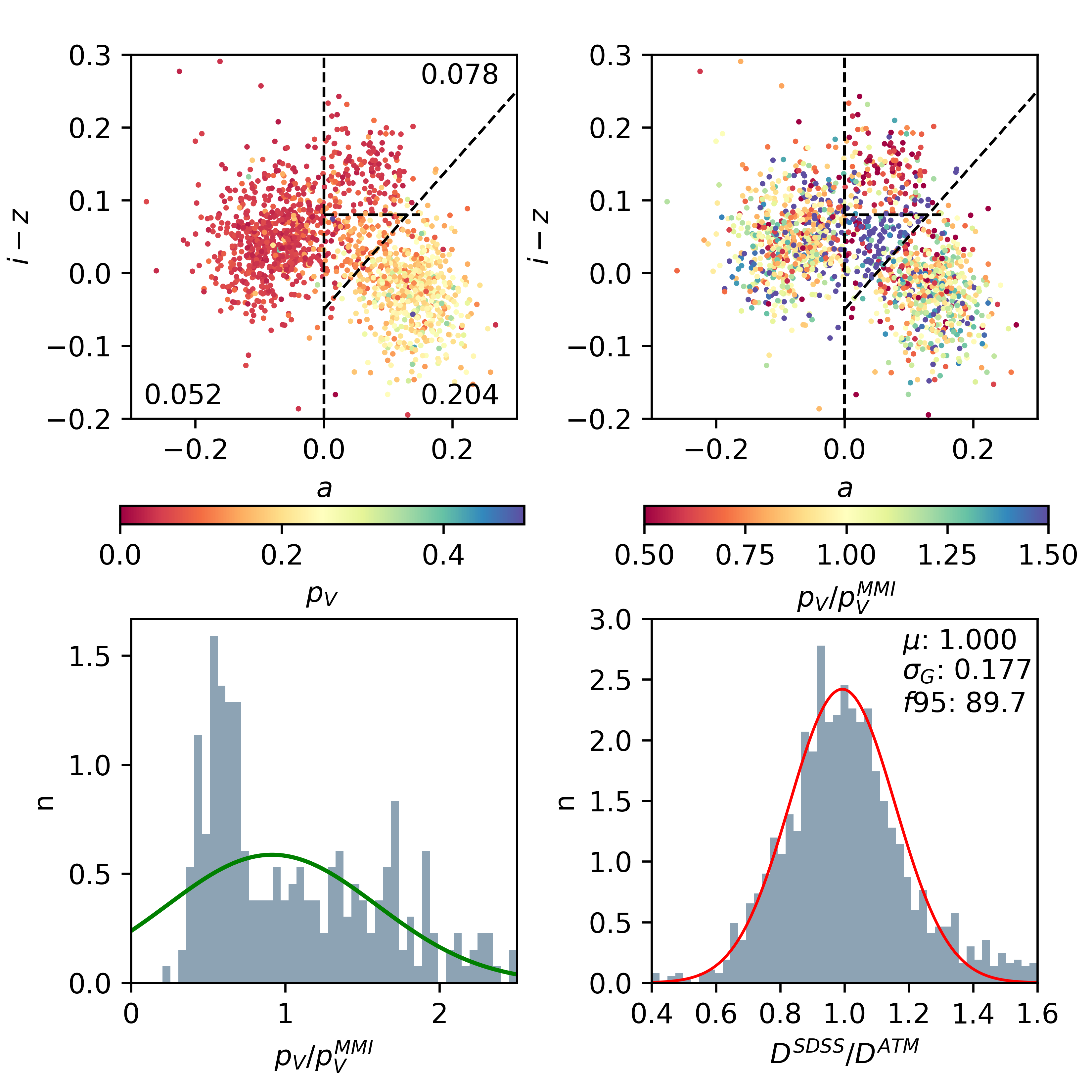}
\caption{\textbf{A revisit of the analysis of the correlations between WISE-based 
model parameters and optical colors measured by SDSS, as discussed in MMI.} 
The top panels show the $a$ vs. $i-z$ SDSS color-color diagram,
where $a$ color is defined as $a=0.89*(g-r) + 0.45*(r-i) - 0.57$ \citep{MOC2001}, for 1,557 
asteroids from SDSS MOC4 catalog that also have WISE-based diameters and
IR albedo estimated with \atm\ (MMI). The symbols in the top left panel are color-coded 
by WISE-based V-band albedo $p_V$, obtained using WISE-based diameter and 
SDSS-based absolute magnitude. The vertical dashed line shows the separation between 
C and S taxonomic classes from \cite{MOC2001}, while the diagonal dashed line is a 
separator between low-albedo (dominated by D type) and high-albedo (dominated by S type) 
objects derived by MMI. The horizontal dashed line is introduced here to improve the MMI results. 
The numbers in the top left panel show the median values of $p_V$, used in a color-based 
model for $p_V$ developed by MMI. The symbols in the top right panel are color-coded 
by the ratio of data-based $p_V$ and this color-based model $p_V$. 
The histogram in the bottom left panel shows the distribution of data/model $p_V$ ratio
for asteroids falling in the triangular region with the median albedo of 0.078 (MMI's
region 2). The dispersion of this ratio is much larger for region 2 (80\%) than for the
other two regions ($\sim$30\%), and motivates separation of region 2 into two 
sub-regions separated by the horizontal dashed line in the two top panels, as proposed 
here. The green line in the bottom left panel is the best-fit single Gaussian. The histogram 
in the bottom right panel shows the distribution of the data/model $p_V$ ratio for 
the MMI model.  The mean ($\mu$), robust standard deviation ($\sigma_G$) and 
the fraction of sample within 2$\sigma_G$ from the median ($f_{95}$, 95\% for normal
distribution) show that SDSS-based diameters match WISE-based best-fit \atm\ values 
with a scatter of 18\% and a reasonably Gaussian distribution.  The line shows the 
corresponding normal (Gaussian) distribution, $N(\mu, \sigma_G)$.
This figure was generated using
\href{https://github.com/ivezicV/2share/blob/master/AsteroidPaper/analyzeSDSSWISE_IIpaper.ipynb}{this python notebook}.\label{fig:MMI}}
\end{figure}

In our analysis, we used the dataset constructed and defined in MMI, 
and used to produce their Figure 15. They matched a sample of 7,359 best-observed asteroids from the NEOWISE dataset
(see their Section 3.1 for detailed selection criteria) to asteroids with optical observations 
listed in the Sloan Digital Sky Survey Moving Object Catalog (hereafter 
SDSSMOC; \citealt{MOC2001, MOC2002, 2002AJ....124.1776J, MOC2008}). 
The 4th SDSSMOC release\footnote{The 4th Release of SDSSMOC is
available from \url{http://faculty.washington.edu/ivezic/sdssmoc/sdssmoc.html}} 
lists astrometric and photometric data for 471,569 detections of moving objects
observed by SDSS prior to March 2007. Of those, 220,101 observations are
linked to 104,449 unique objects with orbits. 
A match between 7,359 objects with high-quality WISE-based sizes estimated 
with \atm, and objects listed in the SDSSMOC catalog yields 1,574 objects.
After rejecting an additional 17 objects with outlying colors, we analyze here
a sample of 1,557 objects. 

Given an estimate for asteroid size based on WISE data, $D$, the visual albedo
$p_V$ is computed for each object as 
\eq{
\label{eq:pVD}
          p_V =  \left( {1329\, {\rm km} \over D} \right)^2  \, 10^{-0.4 H}. 
}
Following MMI, we use optical absolute magnitude, $H$, based on SDSS measurements
to estimate $p_V$. As mentioned in Section~1, the precision of size
estimates based on WISE data varies from 2-3\% for the brightest and best observed 
objects to about 10\% for fainter objects. We assume here that ``typical'' precision
is 5\%. We assume that $H$ uncertainty is 0.05 mag, based on the analysis from
\cite{MOC2008}. This uncertainty is larger than typical SDSS photometric accuracy 
for asteroid observations because of the photometric scatter due to rotation. 
The resulting $p_V$ uncertainty is about 11\% and dominated by $D$ uncertainties. 
We use and further discuss this estimate in Section~\ref{sec:XD}. 

Given these WISE-based estimates of $p_V$, below we use various statistical models to 
map the variation of $p_V$ with SDSS colors and estimate the model $p_V^{model}$, 
which is only a function of SDSS colors. With this model and SDSS data, SDSS-based
size can be estimated by transforming Eq.~\ref{eq:pVD} into
\eq{
\label{eq:opticalD}
            D = 1329\, {\rm km} \, {10^{-0.2 H} \over \sqrt{p_V^{model}}}. 
}
It is straightforward to show that the relevant metric for assessing the contribution of
the $p_V$-color mapping to uncertainty of optical size estimates is simply 
\eq{
\label{eq:Drat}
            \delta_D  = \left( {p_V \over p_V^{model}} \right)^{1/2}.
}
 
It is possible to construct a larger sample of asteroids that have both WISE-based
sizes and data in the SDSSMOC catalog by relaxing the constraint on the quality 
of WISE detections. If all objects with WISE-based sizes are used (dominated by objects 
detected only in the WISE W3 band, see MMI for details), there  are 53,634
matches in the SDSSMOC catalog (42,333 objects with reliable SDSS colors and 
$W3<10$). Single-band W3-based size estimates have an intrinsic precision of about 
10\% relative to the bright high-quality subsample of 7,359 objects with detections in 
all four WISE bands (MMI). Because their size uncertainties 
are larger, the resulting $p_V$ uncertainties are larger, too, and estimated to be about
20\%. We are interested here in high-fidelity calibration of  the mapping between 
optical colors and $p_V$. We opted for the smaller sample because its $p_V$ estimates
are twice as precise as for the larger sample. We did, however, use the larger sample
to verify a few conclusions based on the smaller more accurate sample, as discussed
further below (see Section~\ref{sec:XD}).

\subsection{Constant-albedo models \label{sec:cam}} 

In order to map optical SDSS colors to $p_V$, MMI defined three regions in the 
$i-z$ vs. $a$ color-color diagram (see the top left panel in Figure~\ref{fig:MMI}). 
The asteroid $a$ color corresponds to the first principal axis in the $r-i$ vs. 
$g-r$ color-color diagram and it is defined as \citep{MOC2001},
\eq{
            a = 0.89*(g-r) + 0.45*(r-i) - 0.57. 
}
This color separates the two dominant asteroid taxonomic classes (low-albedo
C type with $a<0$ and high-albedo S type with $a>0$). 
MMI assigned a single value of $p_V$ to each region, that was determined
as the median value of $p_V$ for all asteroids in the corresponding region. 
Asteroid sizes estimated with this simple method agree with WISE-based
sizes with a scatter\footnote{We use robust estimator of standard deviation
computed as $\sigma_G = 0.741*(q_{75}-q_{25})$, where $q_{25}$ and $q_{75}$
are the 25\% and 75\% quantiles, and the normalization factor 0.741 assures
that $\sigma_G$ is equal to standard deviation for normal (Gaussian) distribution.} 
of $\sigma_G=18\%$, as shown in the bottom right panel in Figure~\ref{fig:MMI}. 
The same panel also shows that the fraction of points within $2*\sigma_G$ 
from the median is $f_{95}=89.5\%$, which is close to the value of 95\% 
expected for a normal (Gaussian) distribution. We note that MMI reported 
$\sigma_G=17\%$ -- our result is larger because we did not use any 
outlier rejection scheme when estimating $\sigma_G$. 
 
Our analysis of MMI model residuals suggests that this performance can be
improved. As illustrated in the top right and bottom left panels in Figure~\ref{fig:MMI},
the central wedge-shaped region in the $i-z$ vs. $a$ color-color diagram 
contains at least two types of asteroids distinguished by their different albedo
distributions. The histogram of the data/model ratio, $p_V/p_V^{MMI}$, in the 
bottom left panel shows only asteroids from that region, and it is much wider 
than for the other two regions ($\sigma_G = 80\%$ vs. $\sigma_G = 30\%$, 
as indicated in Table 3 from MMI).  

\begin{figure}[th]
\centering
\includegraphics[width=1.0\textwidth, keepaspectratio]{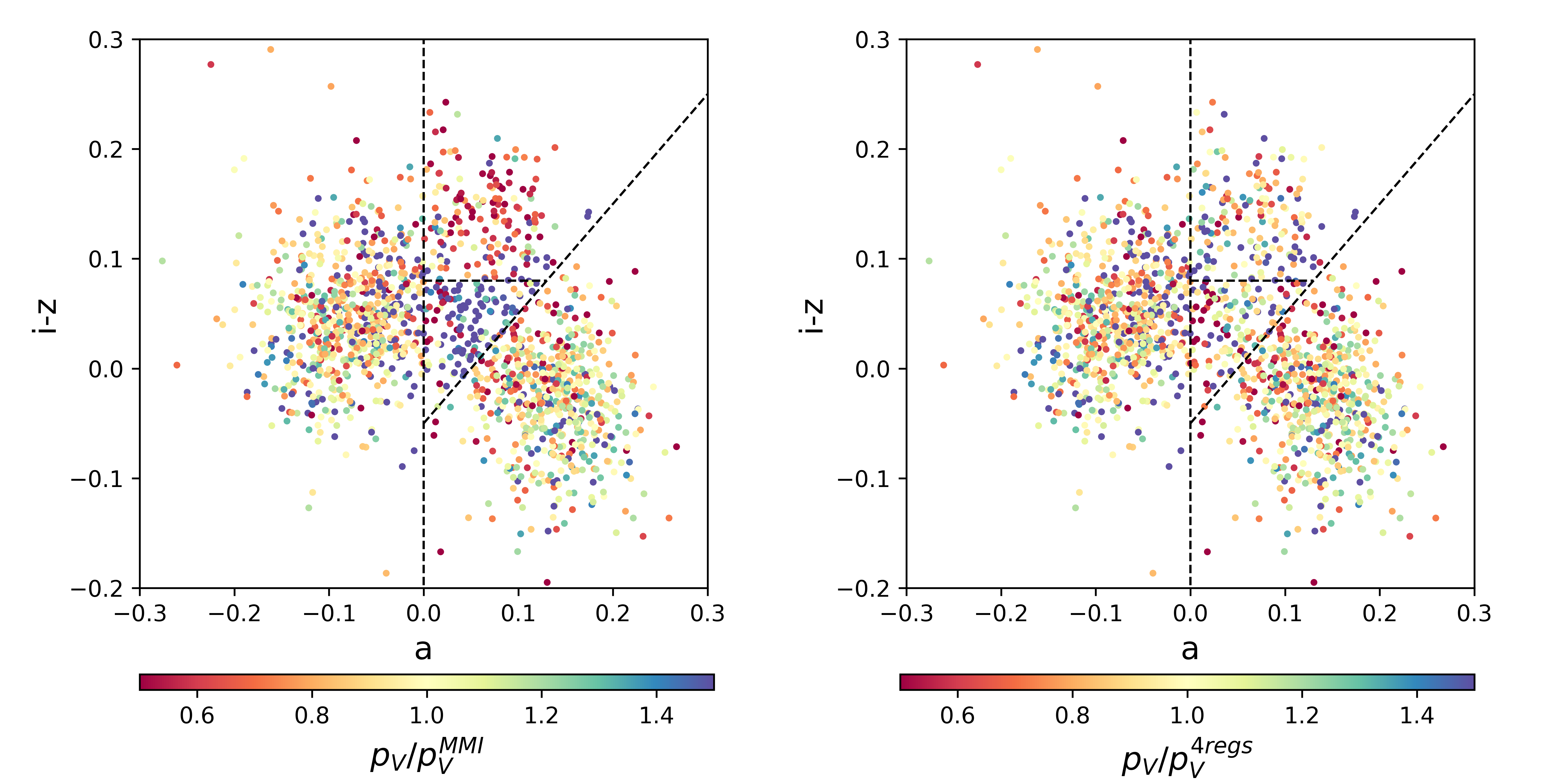}
\caption{\textbf{A comparison of the two methods for mapping optical colors to albedo $p_V$.} 
The symbols in the left panel are color-coded by the ratio of data-based $p_V$ and the color-based 
constant-albedo model $p_V$ developed by \cite{MMI}; the values around unity correspond to good 
matching. The right panel shows the analogous performance of our 
color-based constant-albedo model for $p_V$ that uses an additional region, corresponding to the triangle in the center. 
The robust standard deviation for the ratio of implied size estimates and WISE-based size estimates
is 18\% for the left panel and 16\% for the right panel. This figure was generated using
\href{https://github.com/ivezicV/2share/blob/master/AsteroidPaper/analyzeSDSSWISE_IIpaper.ipynb}{this python notebook}.\label{fig:modelComparisonMMI}}
\end{figure}

Motivated by these findings and the variation of the $p_V/p_V^{MMI}$ ratio with 
color, we split the wedge-shape region into two regions at $i-z = 0.08$. 
We varied the placement of this boundary by $\pm$0.05 mag in steps of 0.01 mag. 
The optimal value of 0.08 could be changed by up to about 0.02 mag without 
appreciable change of the model performance. 
Instead of adopting the median value $p_V = 0.073$ for all objects, we now adopt 
the median value $p_V=0.053$ for objects with $i-z > 0.08$ and $p_V=0.118$ for 
$i-z < 0.08$. 
This new four-region model produces asteroid sizes that agree with WISE-based
sizes with a scatter of $\sigma_G=16\%$.  A comparison of the data/model
ratio, $p_V/p_V^{MMI}$ and $p_V/p_V^{4reg}$ for the original and new methods,
in the $i-z$ vs. $a$ color-color diagram in Figure~\ref{fig:modelComparisonMMI} 
shows marked improvement.

A variety of published data and analysis \citep{MOC2001, 2010A&A...510A..43C, 2012ApJ...745....7M, 2013Icar..226..723D}
suggest that the region with $i-z > 0.08$ and the median $p_V=0.053$ is dominated by D type asteroids. This region also
stands out when the ratio of infrared and optical albedo is considered \citep[see Fig. 6 in][]{2012ApJ...745....7M}. 
The middle triangular region with $i-z < 0.08$ and the median $p_V=0.118$ seems to contain a mixture of L and X
types from the Bus-DeMeo taxonomy. It is noteworthy that \cite{2013Icar..226..723D} set the boundary between 
D and L regions at $i-z=0.085$, in excellent agreement with our result. 

\begin{figure}[th]
\centering
\includegraphics[width=1.0\textwidth, keepaspectratio]{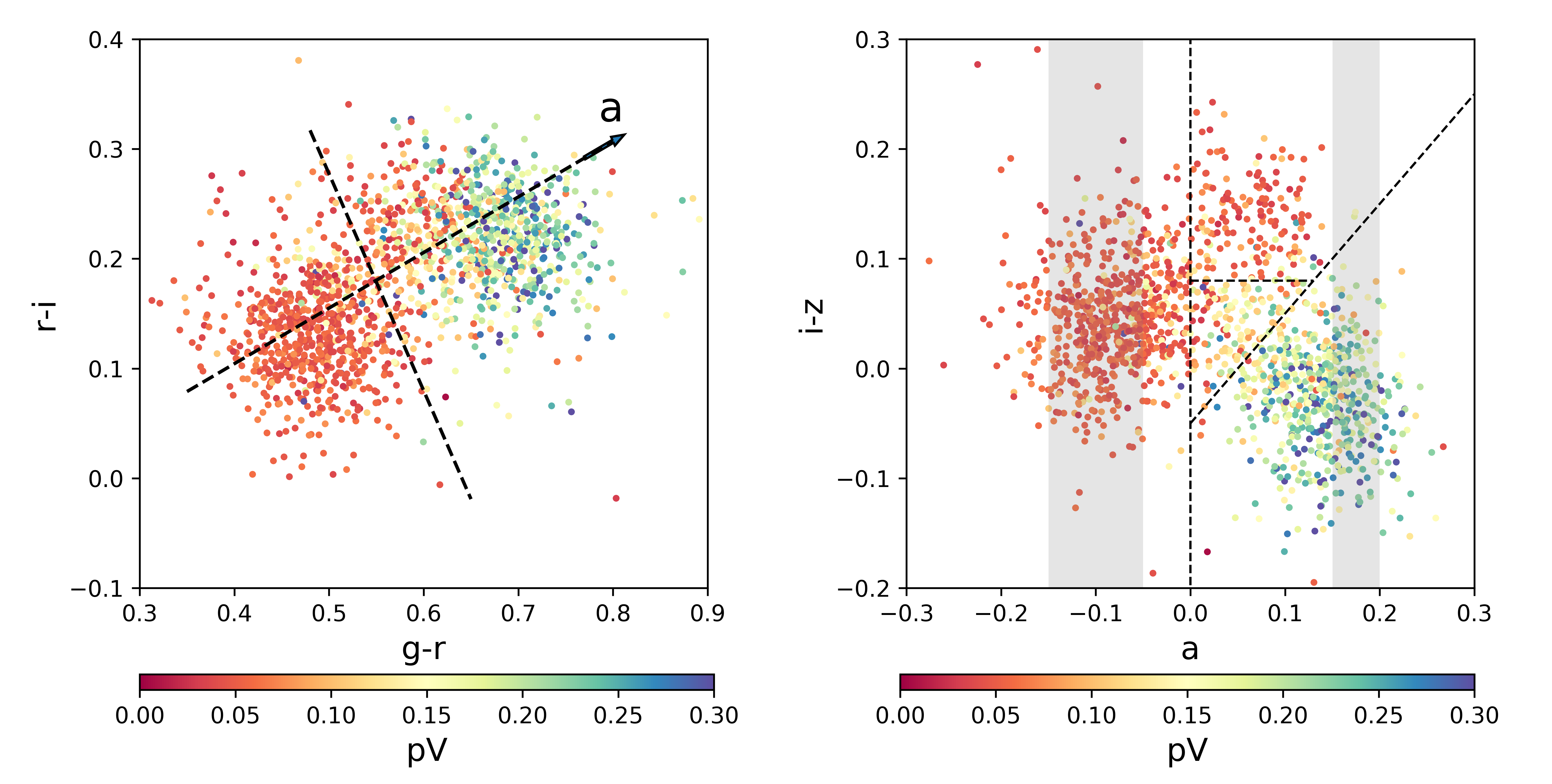}
\caption{\textbf{The SDSS color-color distributions for 1,557 asteroids from SDSS 
MOC4 catalog that also have WISE-based albedo estimates.}  The symbols are 
color-coded by the values of albedo $p_V$. The dashed lines in the left panel show
the principal axes that define the $a$ color. The right panel is analogous to the top
left panel in Figure~\ref{fig:MMI} except for a different $p_V$ scale, as marked at the
bottom. The dashed lines in the right panel are described in the caption of 
Figure~\ref{fig:MMI}. The two gray rectangles mark the $a$ color range used
to select clean subsamples of C type (left) and S type (right) asteroids (for discussion,
see Section~\ref{sec:XD}). Note that the variation of $p_V$ with colors is more pronounced 
in the right panel. This figure was generated using
\href{https://github.com/ivezicV/2share/blob/master/AsteroidPaper/analyzeSDSSWISE_IIpaper.ipynb}{this python notebook}.\label{fig:allcolors}}
\end{figure}

Given the improved performance, it is prudent to ask whether more regions
in the $i-z$ vs. $a$ color-color diagram would yield further improvements,
as well as if additional colors would help. The right panel in Figure~\ref{fig:allcolors} 
shows albedo using a harder color map stretch than in Figure~\ref{fig:MMI}. 
It doesn't seem that another constant-albedo region can be easily defined in
addition to the existing four. As shown in the left panel, the same conclusion 
is valid for the $r-i$ vs. $g-r$ color-color diagram. We reinforce these conclusions
in Section~\ref{sec:GMM} using machine learning tools.

\subsection{Nearest Neighbors method\label{sec:NN}}

Instead of adopting a single albedo value for an entire fixed color region,
we apply the Nearest Neighbors method to assign a different albedo to 
each individual object. Such methods are successfully used in the context
of photometric redshifts of galaxies \citep{2018AJ....155....1G}. We 
investigated the performance as a function of the number of nearest 
neighbors (in the range 3 to 20), and several statistics used to assign the $p_V$ value: 
the mean, the median, and the weighted mean, with the weights 
proportional to $1/\delta_i^2$, where $\delta_i$ is a four-dimensional Euclidean 
distance to the $i^{th}$ neighbor, in the space spanned by SDSS colors. 

As with constant-albedo models, we use the robust standard deviation
for size matching as the relevant performance metric. We find that all three
statistics behave similarly, and that the metric changes little as the 
number of nearest neighbors is varied from 3 to 20, with optimal
values in the range 10 to 15. The WISE-based 
sizes can be reproduced within 16-17\%, thus closely matching the 
performance of the constant-albedo models.  

The analysis of residuals suggests that the performance of the Nearest 
Neighbors method is limited by the sample sparseness. It appears that
at least 10 neighbors are required to reduce the noise due to local $p_V$ scatter; 
however, at the distribution edges and low-density regions quite a large 
color space is spanned to get that many neighbors. We demonstrate in
Section~\ref{sec:XD} that the performance of this method can be
further improved with a larger, model-generated, sample of 500,000 
objects.

\subsection{Gaussian Mixture models \label{sec:GMM}} 

The analysis presented in the preceding section indicated that four regions are required
and sufficient to
capture the variation of albedo $p_V$ with SDSS colors. Here we use a Gaussian Mixture
model and Bayesian model selection to investigate the statistically optimal number of 
Gaussian components in the three-dimensional space spanned by SDSS colors 
$g-r$, $r-i$ and $i-z$ (we do not use the $u-g$ color because it has much 
larger errors than the other three colors for the majority of objects). 

Gaussian Mixture models describe a distribution of data points in the multi-dimensional
space with a sum of multi-variate Gaussian distributions; for more details see
Section 6.3.1 in \cite{zeljkoBook}. We adapted the astroML 
code\footnote{Available from https://www.astroml.org/book\_figures/chapter10/fig\_LINEAR\_clustering.html} 
used to produce figure 10.20 from that book. 

\begin{figure}[th]
\centering
\includegraphics[width=1.0\textwidth, keepaspectratio]{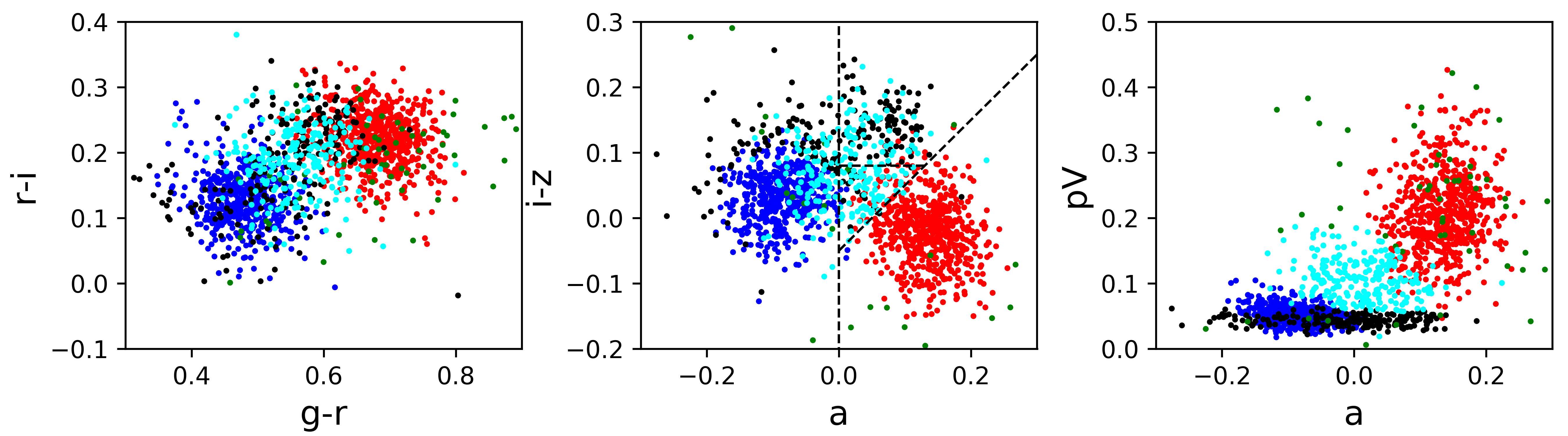}
\caption{\textbf{Gaussian Mixture Model (GMM) applied to the three-dimensional 
space spanned by the $a$ and $i-z$ SDSS colors and WISE-based albedo $p_V$.} 
The GMM class assigned to 1,557 data points is illustrated as color-coded symbols.  
The best model, with the lowest value of BIC (Bayesian Information Criterion), has five
components (shown as blue, red, cyan, black, and green) but the four most dominant 
components include as many as 96\% of all data points. The dashed lines in the middle 
panel show the same regions as in Figure~\ref{fig:MMI}. This figure was generated using
\href{https://github.com/ivezicV/2share/blob/master/AsteroidPaper/analyzeSDSSWISE_IIpaper.ipynb}{this python notebook}.\label{fig:GMMcolors}}
\end{figure}

We run the code with the number of components running from 1 to 20. The statistically
optimal number of components was determined using  the Bayesian Information
Criterion (BIC), for details see Section 5.4 in \cite{zeljkoBook}. The optimal number
of components was five, with 96\% of all points classified as belonging to
one of the four largest components/classes. The resulting model is illustrated in 
Figure~\ref{fig:GMMcolors}. 

The main conclusion of applying Gaussian Mixture model is that the four 
manually defined color regions are supported by this unsupervised procedure. 
In particular, the method autonomously re-discovered that the wedge-shaped region from MMI
includes two populations with very different albedo distributions (although 
the albedo was not used to define the color-based classes). 
 
Given this model, a sample of an arbitrary size could be cloned (generated) and used
as input to the Nearest Neighbor method to test the impact of the dataset sparseness.
We will do such a test, but using a more sophisticated  Gaussian Mixture model described
next.  

\subsection{Extreme Deconvolution model \label{sec:XD}}

The Gaussian Mixture model used in the preceding section does not account
for measurement errors. A more sophisticated version capable of treating
Gaussian errors is known in astronomy as the Extreme Deconvolution (XD)
method\footnote{See also https://github.com/jobovy/extreme-deconvolution}
\citep{2011AnApS...5.1657B}. We used an implementation described in Section
6.3.3 from \cite{zeljkoBook} and adapted the code\footnote{Available from
https://www.astroml.org/book\_figures/chapter6/fig\_XD\_example.html}
used to produce figure 6.11 from that book. 

We apply the XD method to the 4-dimensional space spanned by the SDSS
$g-r$, $r-i$ and $i-z$ colors, and the WISE-based albedo $p_V$. We aim
to capture details in the correlation between $p_V$ and colors and increase 
the number of components from five suggested by GMM to ten. This increased
model flexibility can also account for non-Gaussian shapes of underlying 
clusters. Given the large sample  size, there is no danger of overfitting with that 
many model components: although there are a few dozen free parameters for the 
ten-component XD model, there are many more constraints (objects in the sample). 
We demonstrate further below that the ten-component model appears sufficiently
flexible to capture details in the data. 

A crucial feature of the XD method is accounting for measurement uncertainties 
(errors). The SDSS photometric measurements are obtained in the order $r–i–u–z–g$, 
and the elapsed time between the first ($r$) and last ($g$) measurement is only
about 5 minutes.  The impact of rotational variability on color uncertainties
is thus minimal \citep{2004MNRAS.348..987S}. The typical color uncertainties
for the relatively bright ($r<20$) sample used here are 0.02-0.03 mag. The $p_V$ 
measurement uncertainties were already discussed in Section~\ref{sec:dataset} and 
are estimated to be 11\%. In the XD analysis below we assume that all objects
have this value of measurement uncertainty for $p_V$. 
 
This assumption can be tested by comparing the $p_V$ scatter for subsamples
with uniform taxonomy. As discussed in detail by \cite{2010A&A...510A..43C, 2012ApJ...745....7M, 2013Icar..226..723D}, 
SDSS colors can be used to define subsamples with uniform spectroscopically confirmed
taxonomic classes. Here we analyzed two color-subsamples that correspond to the
most numerous C and S taxonomic types. We selected a C subsample by
a simple cut $-0.15 < a < -0.05$ (see the right panel in Figure~\ref{fig:allcolors}) and 
measured a robust standard deviation for $p_V$ of 26\% around the median value. 
An S subsample, selected by $0.15 < a < 0.20$ shows a similar scatter of 26.5\%. 
Given the measurement uncertainty of 11\%, this measured scatter implies that  
intrinsic (astrophysical) $p_V$ scatter is about 24\% for both subsamples. We also 
used the full sample described in Section~\ref{sec:dataset}, which has $p_V$ uncertainties
of about 20\%.  The measured $p_V$ scatter for both C and S subsamples
is about 30\%, which implies an intrinsic $p_V$ scatter of 22\%, in good agrement 
with the high-quality sample. This internal consistency test supports our 
assumption that all objects have $p_V$ measurement uncertainty of 11\%.

\begin{figure}[th]
\centering
\includegraphics[width=1.0\textwidth, keepaspectratio]{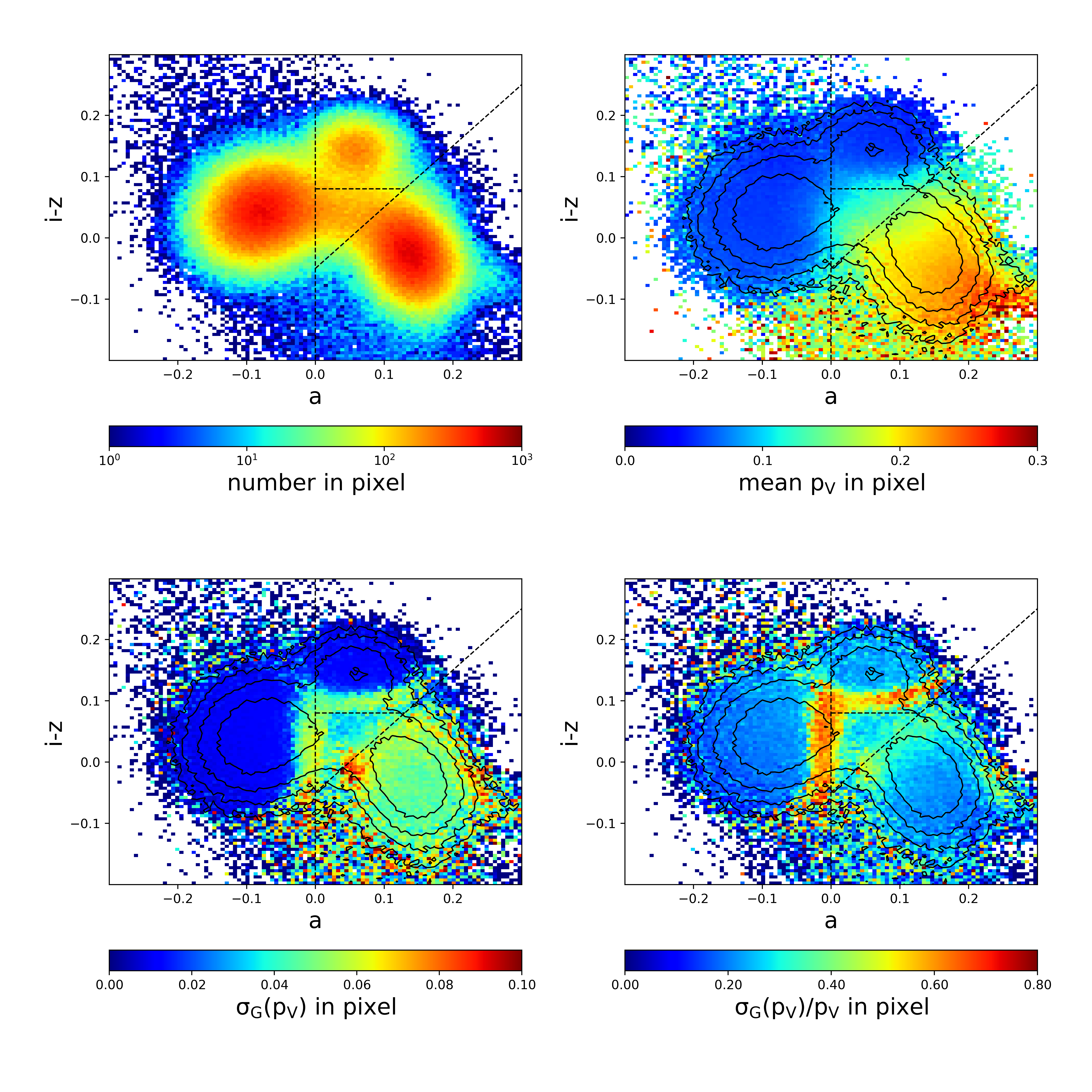}
\caption{\textbf{The color and albedo distributions of a cloned sample with 
500,000 points generated using Extreme Deconvolution model.} The top left panel shows
the counts (two-dimensional histogram) on logarithmic scale. The top right  panel shows the same counts as contours, 
and the mean $p_V$ values are color-coded according to the legend below the panel.
The bottom left panel is analogous to the top right panel, except that robust standard deviation
of $p_V$ is shown instead of the mean values. The bottom right panel shows the ratio
of the bottom left and top right panels (that is, the $p_V$ scatter per pixel normalized by the mean $p_V$ value).
Note that the low-albedo regions, that also have small scatter, are  well defined in color-color space. 
As the bottom right panel illustrates, the relative $p_V$ scatter is quite uniform, except at the
boundary of the triangular region and the low-albedo regions. This figure was generated using
\href{https://github.com/ivezicV/2share/blob/master/AsteroidPaper/analyzeSDSSWISE_IIpaper.ipynb}{this python notebook}.\label{fig:XDdistribution}}
\end{figure}

Figure~\ref{fig:XDdistribution} illustrates the XD model. After obtaining
the best-fit parameters for a 10-component Gaussian mixture, we cloned
a sample of 500,000 points. Their distribution in the $i-z$ vs. $a$ color-color
diagram, as well as the variation of the mean $p_V$ and its dispersion in
the same diagram, is shown in Figure~\ref{fig:XDdistribution}. Although
the model is based on 10 components, there are only about 4-5 clearly 
distinguished components, in agreement with the BIC analysis from 
Section~\ref{sec:GMM}. The behavior of the mean $p_V$ is remarkably
consistent with our constant-albedo model discussion in Section~\ref{sec:cam}. 
In particular, the triangular fourth region that we introduced is clearly
visible in the XD results, although no such prior was used by the XD method. 
The triangular shape also stands out in the dispersion diagram (bottom left panel),
where the transition from one nearly-uniform $p_V$ region to its neighbor
is reflected in increased dispersion. As illustrated in the bottom right panel,  
the relative dispersions (dispersion divided by the correspoding mean $p_V$ 
values) for the blue (C type) and green (S type) regions in the bottom left 
panel correspond to the same relative variation of about 24\%. 
The intrinsic scatter for $p_V$ estimated by the XD method for the full sample
is 26\%, in reasonable agreement with the smaller high-quality sample
(but perhaps indicating that the $p_V$ measurement uncertainty for the full 
sample is a bit larger than 20\%). 

Given the best-fit XD model, we can treat the $p_V$ estimation as a 
case of missing data, and use the three colors to estimate the expectation
value for $p_V$  and its uncertainty. One way to assign $p_V$ is to use the best-fit
parameters of the ten multi-variate Gaussians. Another, simpler, way is to use 
the nearest neighbor method discussed in Section~\ref{sec:NN}. Given the very 
large sample, we can use a relatively large number of nearest neighbors
and investigate the impact of dataset sparseness. 

We found that the best results are obtained with about 100 nearest neighbors:
the XD model produces asteroid sizes that agree with WISE-based sizes with 
a scatter of $\sigma_G=15\%$. With a much smaller number of nearest neighbors,
say 20 or fewer neighbors, the intrinsic noise in $p_V$ is not sufficiently
suppressed. This result explains why the nearest neighbor method applied 
to the dataset itself did not achieve as accurate performance as the constant-albedo
model (see Section~\ref{sec:NN}). For example, the same fraction of sources, 
100 in 500,000, and thus approximately the same distance in the multi-dimensional 
color space, corresponds to only about 3 sources in the dataset, which is insufficient 
to supress the intrinsic noise in $p_V$. On the other hand, going to a much larger number
of nearest neighbors will eventually be counter-productive because the intrinsic variation
of $p_V$ with colors would be ``blurred'' (smoothed out, or  ``washed out'') . For the same 
reason, one cannot use a number of nearest neighbors much larger than 3 with the sparse 
dataset analyzed here.

\begin{figure}[th]
\centering
\includegraphics[width=1.0\textwidth, keepaspectratio]{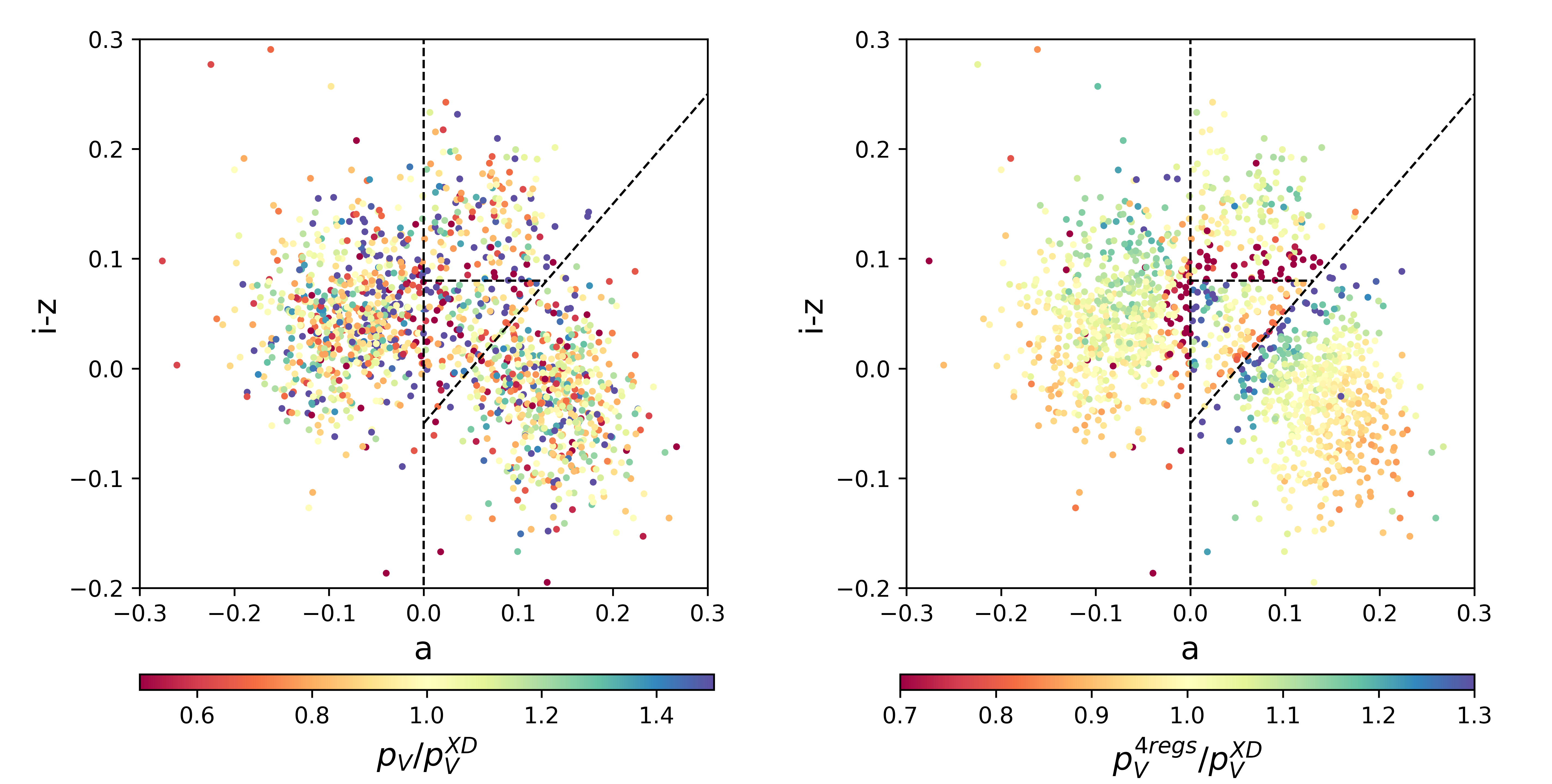}
\caption{
\textbf{A comparison of the two methods for mapping optical colors to albedo $p_V$.} 
The symbols in the left panel are color-coded by the ratio of data-based $p_V$ and the 
color-based model $p_V$ that relies on the nearest neighbor method, with 100 neighbors, 
applied to the Extreme Deconvolution sample (XD model) shown in Figure~\ref{fig:XDdistribution}. 
The robust standard deviation for the ratio of implied size estimates based on the XD
model is 15\%.  For analogous plot with the constant-albedo model results, see the right 
panel in Figure~\ref{fig:modelComparisonMMI}. The right panel shows the model $p_V$ ratio 
for the constant-albedo model developed here and the XD model (note the different color 
scales for the two panels). This figure was generated using
\href{https://github.com/ivezicV/2share/blob/master/AsteroidPaper/analyzeSDSSWISE_IIpaper.ipynb}{this python notebook}.\label{fig:modelcomparisonXD}}
\end{figure}

\begin{figure}[t]
\centering
\includegraphics[width=1.0\textwidth, keepaspectratio]{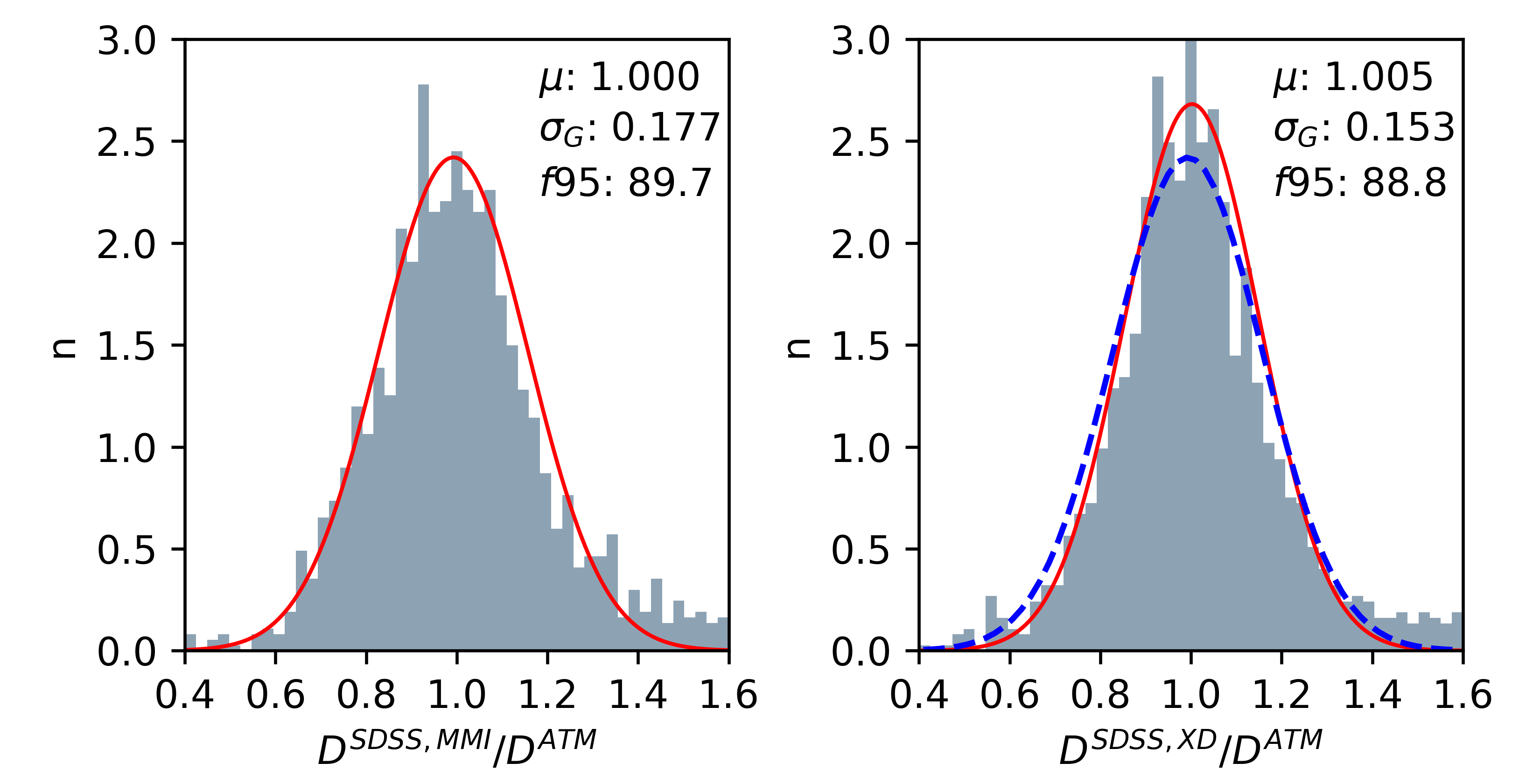}
\caption{
\textbf{A comparison of the MMI method (left panel, same as the bottom right panel in 
Figure~\ref{fig:MMI}) and the XD method (right panel) developed here.}  The dashed
line in the right panel is the same as the solid line in the left panel and illustrates the 
improved performance of the XD method.  
This figure was generated using
\href{https://github.com/ivezicV/2share/blob/master/AsteroidPaper/analyzeSDSSWISE_IIpaper.ipynb}{this python notebook}.\label{fig:MMIvsXDhist}}
\end{figure}

A comparison of residuals for the XD model and the constant-albedo four-regions model 
is illustrated in Figure~\ref{fig:modelcomparisonXD}. There are no strong correlations 
between residuals and colors for the XD method, as shown in the left panel. The right panel
shows the variation of the residuals ratio with colors. The XD method provides improvements
close to the regions boundaries, and also accounts for weak gradients visible for both 
C and S type asteroids. A direct comparison between the MMI and XD performance is 
illustrated in Figure~\ref{fig:MMIvsXDhist}.

\subsection{Extreme Deconvolution model: implications for LSST  \label{sec:XDLSST}}

As described in the preceding section, given the training sample with three colors and $p_V$, 
including their uncertainties, a 10-component Gaussian mixture is fit to this four-dimensional 
distribution.  Using the best-fit model, a large sample of arbitrary size is cloned and used for 
assigning the $p_V$ expectation value and its uncertainty (here with the nearest neighbor method). 
In the cloning step, the intrinsic model colors are convolved with measurement uncertainties 
corresponding to the dataset to which the method is applied. 

Therefore, the same best-fit XD model developed here can be used to predict the precision
of size estimated with LSST data, if the correct color uncertainties expected for LSST are used in the cloning
step. Unlike SDSS measurements, LSST multi-band measurements will not be simultaneous.
For this reason, one needs to account for rotational variability. The root-mean-square 
uncertainty of a single photometric asteroid measurement due to rotation, both in optical and infrared,  
is about $\sigma_{rot} = 0.15$ mag \citep{2000Icar..148...12P, 2008Icar..196..135S, MMI}. 

LSST will obtain about 200-300 photometric measurements per asteroid for over 5 million 
objects during its 10 year survey (see Section 5 in \citealt{2019ApJ...873..111I}). 
Given the small color variability, all bands can be combined together in order to estimate
the rotational period and fit light curve (for a detailed discussion of multi-band periodograms, 
see \citealt{2015ApJ...812...18V}). If a good fit can be found, the best resulting color accuracy
should be of the order 0.01 mag for sufficiently bright asteroids. If, on the other hand,
simple light curve averaging is invoked, then the expected color error is
\eq{
         \sigma_{c} = \left(2\, {\sigma_{phot}^2 + \sigma_{rot}^2  \over N_{obs}}  \right)^{1/2},
}
where the factor of 2 accounts for assumed uncorrelated observations in two bands, 
the photometric error, $\sigma_{phot}$, depends on the photometric signal-to-noise ratio,
and $N_{obs}$ is the number of observations used in the averaging (250 on average over 
10 years and, assuming uniform observing pattern, 25 per year). For an asteroid that is 
always at the detection limit so that $\sigma_{phot}=0.2$ mag, the resulting color uncertainty 
is 0.02 mag after 10 years and 0.07 mag after 1 year of LSST data. Given these estimates, we
investigated the performance of the XD method using two assumptions for color errors: 
an optimistic 0.01 mag and a pessimistic 0.05 mag. We note that intrinsic asteroid color variations
due to rotation estimated from repeated SDSS measurements are about 0.03 mag \citep{2004MNRAS.348..987S}. 

The prediction of the XD method is that with 0.01 mag color uncertainty, the accuracy of
predicted sizes would be 14\% (compare to 15\% achieved with SDSS data).  This result
implies that the limiting factor is the correlation of albedo with colors, and its degeneracies,
rather than photometric accuracy. With color uncertainty of 0.05 mag, the accuracy of
predicted sizes would be about 16\%. We also investigated larger color errors and found that 
the mapping from colors to albedo becomes ``blurred'' as 
the color errors increase beyond 0.1 mag, in agreement with expectation given the dynamic
range of asteroid optical colors ($\sim$0.5 mag).

\section{Discussion and conclusions}

We revisited a correlation between SDSS optical colors and optical albedo derived using 
WISE-based size estimates and developed several improved methods to estimate asteroid sizes 
with optical data alone. The best-performing approach uses the so-called Extreme Deconvolution method,
a Gaussian mixture model that accounts for measurement uncertainties, to clone a large sample
statistically consistent with the data, and then assigns the best-fit albedo and its 
uncertainty using nearest neighbors. This method reproduces WISE-based asteroid size estimates
with a scatter of 15\%. This good performance bodes well for future optical asteroid 
surveys, such as the Rubin Observatory Legacy Survey of Space and Time, which will 
deliver such size estimates for over 5 million asteroids. 
 
Optical color-based size estimates calibrated to agree with WISE-based size estimates   
with a precision of 15\% yield the size accuracy in the range 21-25\% after addition 
of WISE-based accuracy, 15-20\%, in quadrature. Therefore, the accuracy of optical 
color-based size estimates is only a factor of 1.3 to 1.4 worse than for IR-based estimates 
when sufficiently accurate multi-band optical photometry and IR-based training sample are 
available. This size estimation accuracy is significantly better than commonly assumed 
for optical data. For example, the recent NAS report ``Finding Hazardous Asteroids Using Infrared 
and Visible Wavelength Telescopes'' \citep{NAP25476}, conservatively assumed that 
optical size estimates are {\it as much as a factor of 4 worse} than infrared-based size 
estimates, which may impact some of the conclusions presented there. 
The much better, about three times, optical size estimation accuracy demonstrated 
here is due to accurate and homogeneous multi-band photometry delivered by SDSS, large 
accurate calibration sample delivered by the WISE survey, and adequate methods for mapping 
colors to albedo discussed here. 

Using the best-fit Extreme Deconvolution model, we estimated that intrinsic $p_V$ 
scatter for the most numerous C and S asteroids selected by colors is about 24\%. This value implies that the 
optical size estimate uncertainty could be ultimately pushed down to $\sim$12\% level, 
if a sufficiently large and accurate calibration sample were available in addition to
optical LSST photometry. 
At least two upcoming infrared missions have potential to obtain such calibration samples:
the Near-Earth Object Surveillance Mission\footnote{https://neocam.ipac.caltech.edu} and SPHEREx\footnote{https://spherex.caltech.edu/index.html}. 
In addition, direct asteroid size measurements are required for robust 
and precise calibration of thermal asteroid models and quantitative estimation
of their intrinsic biases.

\subsection{Data Availability} 

All the analysis presented here can be easily reproduced with our data files
and Python Notebook that are made publicly available from this 
GitHub\footnote{See \url{https://github.com/ivezicV/2share/tree/master/AsteroidPaper}} site.

\subsection{Acknowledgments}

The authors thank Joachim Moeyens, Melissa Graham, Dino Bekte\v{s}evi\'{c}, Siegfried Eggl, Lynne Jones and 
Mario Juri\'{c} from the University of Washington for their comments and expert advice. 
We would also like to thank the two anonymous reviewers for their comments that helped 
improve the presentation of our results.  

\v{Z}. Ivezi\'{c} acknowledges support from the University of Washington College of Arts and Sciences, 
Department of Astronomy, and the DIRAC Institute. The DIRAC Institute is supported through generous 
gifts from the Charles and Lisa Simonyi Fund for Arts and Sciences, and the Washington Research Foundation.  
 
This publication makes use of data products from the Wide-field Infrared Survey Explorer, which is a joint project of the University of California, Los Angeles, and the Jet Propulsion Laboratory/California Institute of Technology, funded by the National Aeronautics and Space Administration.
 
\software{numpy \citep{numpy}, matplotlib \citep{matplotlib}, scipy \citep{scipy}, astropy \citep{astropy-1, astropy-2}, astroML \citep{2012cidu.conf...47V}.}

\bibliographystyle{aasjournal}
\bibliography{ref}{}

\begin{thebibliography}{}
\expandafter\ifx\csname natexlab\endcsname\relax\def\natexlab#1{#1}\fi

\bibitem[{{Astropy Collaboration} {et~al.}(2013){Astropy Collaboration},
  {Robitaille}, {Tollerud}, {Greenfield}, {Droettboom}, {Bray}, {Aldcroft},
  {Davis}, {Ginsburg}, {Price-Whelan}, {Kerzendorf}, {Conley}, {Crighton},
  {Barbary}, {Muna}, {Ferguson}, {Grollier}, {Parikh}, {Nair}, {Unther},
  {Deil}, {Woillez}, {Conseil}, {Kramer}, {Turner}, {Singer}, {Fox}, {Weaver},
  {Zabalza}, {Edwards}, {Azalee Bostroem}, {Burke}, {Casey}, {Crawford},
  {Dencheva}, {Ely}, {Jenness}, {Labrie}, {Lim}, {Pierfederici}, {Pontzen},
  {Ptak}, {Refsdal}, {Servillat}, \& {Streicher}}]{astropy-1}
{Astropy Collaboration}, {Robitaille}, T.~P., {Tollerud}, E.~J., {et~al.} 2013,
  \aap, 558, A33

\bibitem[{{Astropy Collaboration} {et~al.}(2018){Astropy Collaboration},
  {Price-Whelan}, {Sip{\H o}cz}, {G{\"u}nther}, {Lim}, {Crawford}, {Conseil},
  {Shupe}, {Craig}, {Dencheva}, {Ginsburg}, {VanderPlas}, {Bradley},
  {P{\'e}rez-Su{\'a}rez}, {de Val-Borro}, {Aldcroft}, {Cruz}, {Robitaille},
  {Tollerud}, {Ardelean}, {Babej}, {Bach}, {Bachetti}, {Bakanov}, {Bamford},
  {Barentsen}, {Barmby}, {Baumbach}, {Berry}, {Biscani}, {Boquien}, {Bostroem},
  {Bouma}, {Brammer}, {Bray}, {Breytenbach}, {Buddelmeijer}, {Burke},
  {Calderone}, {Cano Rodr{\'{\i}}guez}, {Cara}, {Cardoso}, {Cheedella},
  {Copin}, {Corrales}, {Crichton}, {D'Avella}, {Deil}, {Depagne}, {Dietrich},
  {Donath}, {Droettboom}, {Earl}, {Erben}, {Fabbro}, {Ferreira}, {Finethy},
  {Fox}, {Garrison}, {Gibbons}, {Goldstein}, {Gommers}, {Greco}, {Greenfield},
  {Groener}, {Grollier}, {Hagen}, {Hirst}, {Homeier}, {Horton}, {Hosseinzadeh},
  {Hu}, {Hunkeler}, {Ivezi{\'c}}, {Jain}, {Jenness}, {Kanarek}, {Kendrew},
  {Kern}, {Kerzendorf}, {Khvalko}, {King}, {Kirkby}, {Kulkarni}, {Kumar},
  {Lee}, {Lenz}, {Littlefair}, {Ma}, {Macleod}, {Mastropietro}, {McCully},
  {Montagnac}, {Morris}, {Mueller}, {Mumford}, {Muna}, {Murphy}, {Nelson},
  {Nguyen}, {Ninan}, {N{\"o}the}, {Ogaz}, {Oh}, {Parejko}, {Parley}, {Pascual},
  {Patil}, {Patil}, {Plunkett}, {Prochaska}, {Rastogi}, {Reddy Janga},
  {Sabater}, {Sakurikar}, {Seifert}, {Sherbert}, {Sherwood-Taylor}, {Shih},
  {Sick}, {Silbiger}, {Singanamalla}, {Singer}, {Sladen}, {Sooley},
  {Sornarajah}, {Streicher}, {Teuben}, {Thomas}, {Tremblay}, {Turner},
  {Terr{\'o}n}, {van Kerkwijk}, {de la Vega}, {Watkins}, {Weaver}, {Whitmore},
  {Woillez}, {Zabalza}, \& {Astropy Contributors}}]{astropy-2}
{Astropy Collaboration}, {Price-Whelan}, A.~M., {Sip{\H o}cz}, B.~M., {et~al.}
  2018, \aj, 156, 123

\bibitem[{{Bovy} {et~al.}(2011){Bovy}, {Hogg}, \&
  {Roweis}}]{2011AnApS...5.1657B}
{Bovy}, J., {Hogg}, D.~W., \& {Roweis}, S.~T. 2011, Annals of Applied
  Statistics, 5, 1657

\bibitem[{{Carvano} {et~al.}(2010){Carvano}, {Hasselmann}, {Lazzaro}, \&
  {Moth{\'e}-Diniz}}]{2010A&A...510A..43C}
{Carvano}, J.~M., {Hasselmann}, P.~H., {Lazzaro}, D., \& {Moth{\'e}-Diniz}, T.
  2010, \aap, 510, A43

\bibitem[{{DeMeo} \& {Carry}(2013)}]{2013Icar..226..723D}
{DeMeo}, F.~E., \& {Carry}, B. 2013, \icarus, 226, 723

\bibitem[{{Graham} {et~al.}(2018){Graham}, {Connolly}, {Ivezi{\'c}}, {Schmidt},
  {Jones}, {Juri{\'c}}, {Daniel}, \& {Yoachim}}]{2018AJ....155....1G}
{Graham}, M.~L., {Connolly}, A.~J., {Ivezi{\'c}}, {\v{Z}}., {et~al.} 2018, \aj,
  155, 1

\bibitem[{Hunter(2007)}]{matplotlib}
Hunter, J.~D. 2007, Computing In Science \& Engineering, 9, 90

\bibitem[{{Ivezi{\'c}} {et~al.}(2014){Ivezi{\'c}}, {Connolly}, {VanderPlas}, \&
  {Gray}}]{zeljkoBook}
{Ivezi{\'c}}, {\v Z}., {Connolly}, A.~J., {VanderPlas}, J.~T., \& {Gray}, A.
  2014, {Statistics, Data Mining, and Machine Learning in Astronomy: a
  practical Python guide for the analysis of survey data}, Princeton Series in
  Modern Observational Astronomy (Princeton, NJ: Princeton University Press)

\bibitem[{{Ivezi{\'c}} {et~al.}(2001){Ivezi{\'c}}, {Tabachnik}, {Rafikov},
  {Lupton}, {Quinn}, {Hammergren}, {Eyer}, {Chu}, {Armstrong}, {Fan},
  {Finlator}, {Geballe}, {Gunn}, {Hennessy}, {Knapp}, {Leggett}, {Munn},
  {Pier}, {Rockosi}, {Schneider}, {Strauss}, {Yanny}, {Brinkmann}, {Csabai},
  {Hindsley}, {Kent}, {Lamb}, {Margon}, {McKay}, {Smith}, {Waddel}, {York}, \&
  {SDSS Collaboration}}]{MOC2001}
{Ivezi{\'c}}, {\v Z}., {Tabachnik}, S., {Rafikov}, R., {et~al.} 2001, \aj, 122,
  2749

\bibitem[{{Ivezi{\'c}} {et~al.}(2002){Ivezi{\'c}}, {Lupton}, {Juri{\'c}},
  {Tabachnik}, {Quinn}, {Gunn}, {Knapp}, {Rockosi}, \& {Brinkmann}}]{MOC2002}
{Ivezi{\'c}}, {\v Z}., {Lupton}, R.~H., {Juri{\'c}}, M., {et~al.} 2002, \aj,
  124, 2943

\bibitem[{{Ivezi{\'c}} {et~al.}(2008){Ivezi{\'c}}, {Sesar}, {Juri{\'c}},
  {Bond}, {Dalcanton}, {Rockosi}, {Yanny}, {Newberg}, {Beers}, {Allende
  Prieto}, {Wilhelm}, {Lee}, {Sivarani}, {Norris}, {Bailer-Jones}, {Re
  Fiorentin}, {Schlegel}, {Uomoto}, {Lupton}, {Knapp}, {Gunn}, {Covey}, {Allyn
  Smith}, {Miknaitis}, {Doi}, {Tanaka}, {Fukugita}, {Kent}, {Finkbeiner},
  {Munn}, {Pier}, {Quinn}, {Hawley}, {Anderson}, {Kiuchi}, {Chen}, {Bushong},
  {Sohi}, {Haggard}, {Kimball}, {Barentine}, {Brewington}, {Harvanek},
  {Kleinman}, {Krzesinski}, {Long}, {Nitta}, {Snedden}, {Lee}, {Harris},
  {Brinkmann}, {Schneider}, \& {York}}]{2008ApJ...684..287I}
{Ivezi{\'c}}, {\v{Z}}., {Sesar}, B., {Juri{\'c}}, M., {et~al.} 2008, \apj, 684,
  287

\bibitem[{{Ivezi{\'c}} {et~al.}(2019){Ivezi{\'c}}, {Kahn}, {Tyson}, {Abel},
  {Acosta}, {Allsman}, {Alonso}, {AlSayyad}, {Anderson}, {Andrew}, {Angel},
  {Angeli}, {Ansari}, {Antilogus}, {Araujo}, {Armstrong}, {Arndt}, {Astier},
  {Aubourg}, {Auza}, {Axelrod}, {Bard}, {Barr}, {Barrau}, {Bartlett}, {Bauer},
  {Bauman}, {Baumont}, {Bechtol}, {Bechtol}, {Becker}, {Becla}, {Beldica},
  {Bellavia}, {Bianco}, {Biswas}, {Blanc}, {Blazek}, {Bland ford}, {Bloom},
  {Bogart}, {Bond}, {Booth}, {Borgland}, {Borne}, {Bosch}, {Boutigny},
  {Brackett}, {Bradshaw}, {Brand t}, {Brown}, {Bullock}, {Burchat}, {Burke},
  {Cagnoli}, {Calabrese}, {Callahan}, {Callen}, {Carlin}, {Carlson}, {Chand
  rasekharan}, {Charles-Emerson}, {Chesley}, {Cheu}, {Chiang}, {Chiang},
  {Chirino}, {Chow}, {Ciardi}, {Claver}, {Cohen-Tanugi}, {Cockrum}, {Coles},
  {Connolly}, {Cook}, {Cooray}, {Covey}, {Cribbs}, {Cui}, {Cutri}, {Daly},
  {Daniel}, {Daruich}, {Daubard}, {Daues}, {Dawson}, {Delgado}, {Dellapenna},
  {de Peyster}, {de Val-Borro}, {Digel}, {Doherty}, {Dubois},
  {Dubois-Felsmann}, {Durech}, {Economou}, {Eifler}, {Eracleous}, {Emmons},
  {Fausti Neto}, {Ferguson}, {Figueroa}, {Fisher-Levine}, {Focke}, {Foss},
  {Frank}, {Freemon}, {Gangler}, {Gawiser}, {Geary}, {Gee}, {Geha}, {Gessner},
  {Gibson}, {Gilmore}, {Glanzman}, {Glick}, {Goldina}, {Goldstein}, {Goodenow},
  {Graham}, {Gressler}, {Gris}, {Guy}, {Guyonnet}, {Haller}, {Harris},
  {Hascall}, {Haupt}, {Hernand ez}, {Herrmann}, {Hileman}, {Hoblitt},
  {Hodgson}, {Hogan}, {Howard}, {Huang}, {Huffer}, {Ingraham}, {Innes},
  {Jacoby}, {Jain}, {Jammes}, {Jee}, {Jenness}, {Jernigan}, {Jevremovi{\'c}},
  {Johns}, {Johnson}, {Johnson}, {Jones}, {Juramy-Gilles}, {Juri{\'c}},
  {Kalirai}, {Kallivayalil}, {Kalmbach}, {Kantor}, {Karst}, {Kasliwal},
  {Kelly}, {Kessler}, {Kinnison}, {Kirkby}, {Knox}, {Kotov}, {Krabbendam},
  {Krughoff}, {Kub{\'a}nek}, {Kuczewski}, {Kulkarni}, {Ku}, {Kurita}, {Lage},
  {Lambert}, {Lange}, {Langton}, {Le Guillou}, {Levine}, {Liang}, {Lim},
  {Lintott}, {Long}, {Lopez}, {Lotz}, {Lupton}, {Lust}, {MacArthur}, {Mahabal},
  {Mand elbaum}, {Markiewicz}, {Marsh}, {Marshall}, {Marshall}, {May},
  {McKercher}, {McQueen}, {Meyers}, {Migliore}, {Miller}, {Mills}, {Miraval},
  {Moeyens}, {Moolekamp}, {Monet}, {Moniez}, {Monkewitz}, {Montgomery},
  {Morrison}, {Mueller}, {Muller}, {Mu{\~n}oz Arancibia}, {Neill}, {Newbry},
  {Nief}, {Nomerotski}, {Nordby}, {O'Connor}, {Oliver}, {Olivier}, {Olsen},
  {O'Mullane}, {Ortiz}, {Osier}, {Owen}, {Pain}, {Palecek}, {Parejko},
  {Parsons}, {Pease}, {Peterson}, {Peterson}, {Petravick}, {Libby Petrick},
  {Petry}, {Pierfederici}, {Pietrowicz}, {Pike}, {Pinto}, {Plante}, {Plate},
  {Plutchak}, {Price}, {Prouza}, {Radeka}, {Rajagopal}, {Rasmussen},
  {Regnault}, {Reil}, {Reiss}, {Reuter}, {Ridgway}, {Riot}, {Ritz}, {Robinson},
  {Roby}, {Roodman}, {Rosing}, {Roucelle}, {Rumore}, {Russo}, {Saha},
  {Sassolas}, {Schalk}, {Schellart}, {Schindler}, {Schmidt}, {Schneider},
  {Schneider}, {Schoening}, {Schumacher}, {Schwamb}, {Sebag}, {Selvy},
  {Sembroski}, {Seppala}, {Serio}, {Serrano}, {Shaw}, {Shipsey}, {Sick},
  {Silvestri}, {Slater}, {Smith}, {Smith}, {Sobhani}, {Soldahl},
  {Storrie-Lombardi}, {Stover}, {Strauss}, {Street}, {Stubbs}, {Sullivan},
  {Sweeney}, {Swinbank}, {Szalay}, {Takacs}, {Tether}, {Thaler}, {Thayer},
  {Thomas}, {Thornton}, {Thukral}, {Tice}, {Trilling}, {Turri}, {Van Berg},
  {Vanden Berk}, {Vetter}, {Virieux}, {Vucina}, {Wahl}, {Walkowicz}, {Walsh},
  {Walter}, {Wang}, {Wang}, {Warner}, {Wiecha}, {Willman}, {Winters},
  {Wittman}, {Wolff}, {Wood-Vasey}, {Wu}, {Xin}, {Yoachim}, \&
  {Zhan}}]{2019ApJ...873..111I}
{Ivezi{\'c}}, {\v{Z}}., {Kahn}, S.~M., {Tyson}, J.~A., {et~al.} 2019, \apj,
  873, 111

\bibitem[{Jones {et~al.}(2001--)Jones, Oliphant, Peterson, {et~al.}}]{scipy}
Jones, E., Oliphant, T., Peterson, P., {et~al.} 2001--, {SciPy}: Open source
  scientific tools for {Python}, , , [Online; accessed <today>]

\bibitem[{{Jones} {et~al.}(2018){Jones}, {Slater}, {Moeyens}, {Allen},
  {Axelrod}, {Cook}, {Ivezi{\'c}}, {Juri{\'c}}, {Myers}, \&
  {Petry}}]{2018Icar..303..181J}
{Jones}, R.~L., {Slater}, C.~T., {Moeyens}, J., {et~al.} 2018, \icarus, 303,
  181

\bibitem[{{Juri{\'c}} {et~al.}(2002){Juri{\'c}}, {Ivezi{\'c}}, {Lupton},
  {Quinn}, {Tabachnik}, {Fan}, {Gunn}, {Hennessy}, {Knapp}, {Munn}, {Pier},
  {Rockosi}, {Schneider}, {Brinkmann}, {Csabai}, \&
  {Fukugita}}]{2002AJ....124.1776J}
{Juri{\'c}}, M., {Ivezi{\'c}}, {\v Z}., {Lupton}, R.~H., {et~al.} 2002, \aj,
  124, 1776

\bibitem[{{Mainzer} {et~al.}(2015){Mainzer}, {Usui}, \&
  {Trilling}}]{MainzerReview}
{Mainzer}, A., {Usui}, F., \& {Trilling}, D.~E. 2015, in Asteroids IV, ed.
  P.~Michel, F.~E. DeMeo, \& W.~F. Bottke (Tucson, AZ: University of Arizona
  Press), 89--106

\bibitem[{{Mainzer} {et~al.}(2012){Mainzer}, {Masiero}, {Grav}, {Bauer},
  {Tholen}, {McMillan}, {Wright}, {Spahr}, {Cutri}, {Walker}, {Mo}, {Watkins},
  {Hand }, \& {Maleszewski}}]{2012ApJ...745....7M}
{Mainzer}, A., {Masiero}, J., {Grav}, T., {et~al.} 2012, \apj, 745, 7

\bibitem[{{Mainzer} {et~al.}(2016){Mainzer}, {Bauer}, {Cutri}, {Grav},
  {Kramer}, {Masiero}, {Nugent}, {Sonnett}, {Stevenson}, \&
  {Wright}}]{2016PDSS..247.....M}
{Mainzer}, A.~K., {Bauer}, J.~M., {Cutri}, R.~M., {et~al.} 2016, NASA Planetary
  Data System

\bibitem[{{Masiero} {et~al.}(2018){Masiero}, {Mainzer}, \&
  {Wright}}]{2018AJ.156.62M}
{Masiero}, J.~R., {Mainzer}, A.~K., \& {Wright}, E.~L. 2018, \aj, 156, 62

\bibitem[{{Moeyens} {et~al.}(2020){Moeyens}, {Myhrvold}, \& {Ivezi{\'c}}}]{MMI}
{Moeyens}, J., {Myhrvold}, N., \& {Ivezi{\'c}}, {\v{Z}}. 2020, \icarus, 341,
  113575

\bibitem[{{Mommert} {et~al.}(2018){Mommert}, {Jedicke}, \&
  {Trilling}}]{2018AJ....155...74M}
{Mommert}, M., {Jedicke}, R., \& {Trilling}, D.~E. 2018, \aj, 155, 74

\bibitem[{{Myhrvold}(2018{\natexlab{a}})}]{NM2018b}
{Myhrvold}, N. 2018{\natexlab{a}}, \icarus, 314, 64

\bibitem[{{Myhrvold}(2018{\natexlab{b}})}]{NM2018a}
---. 2018{\natexlab{b}}, \icarus, 303, 91

\bibitem[{{National Academies of Sciences, Engineering, and
  Medicine}(2019)}]{NAP25476}
{National Academies of Sciences, Engineering, and Medicine}. 2019, {Finding
  Hazardous Asteroids Using Infrared and Visible Wavelength Telescopes}
  (Washington, DC: The National Academies Press), doi:10.17226/25476

\bibitem[{Oliphant(2006)}]{numpy}
Oliphant, T.~E. 2006, A guide to NumPy, Vol.~1 (Trelgol Publishing USA)

\bibitem[{{Parker} {et~al.}(2008){Parker}, {Ivezi{\'c}}, {Juri{\'c}}, {Lupton},
  {Sekora}, \& {Kowalski}}]{MOC2008}
{Parker}, A., {Ivezi{\'c}}, {\v Z}., {Juri{\'c}}, M., {et~al.} 2008, \icarus,
  198, 138

\bibitem[{{Pravec} \& {Harris}(2000)}]{2000Icar..148...12P}
{Pravec}, P., \& {Harris}, A.~W. 2000, \icarus, 148, 12

\bibitem[{{Szab{\'o}} {et~al.}(2004){Szab{\'o}}, {Ivezi{\'c}}, {Juri{\'c}},
  {Lupton}, \& {Kiss}}]{2004MNRAS.348..987S}
{Szab{\'o}}, G.~M., {Ivezi{\'c}}, {\v{Z}}., {Juri{\'c}}, M., {Lupton}, R., \&
  {Kiss}, L.~L. 2004, \mnras, 348, 987

\bibitem[{{Szab{\'o}} \& {Kiss}(2008)}]{2008Icar..196..135S}
{Szab{\'o}}, G.~M., \& {Kiss}, L.~L. 2008, \icarus, 196, 135

\bibitem[{{VanderPlas} {et~al.}(2012){VanderPlas}, {Connolly}, {Ivezi{\'c}}, \&
  {Gray}}]{2012cidu.conf...47V}
{VanderPlas}, J., {Connolly}, A.~J., {Ivezi{\'c}}, {\v Z}., \& {Gray}, A. 2012,
  in Proceedings of Conference on Intelligent Data Understanding (CIDU), pp.
  47-54, 2012., 47--54

\bibitem[{{VanderPlas} \& {Ivezi{\'c}}(2015)}]{2015ApJ...812...18V}
{VanderPlas}, J.~T., \& {Ivezi{\'c}}, {\v{Z}}. 2015, \apj, 812, 18

\bibitem[{{Wright}(2007)}]{2007astro.ph..3085W}
{Wright}, E.~L. 2007, arXiv e-prints, astro

\bibitem[{{Wright} {et~al.}(2010){Wright}, {Eisenhardt}, {Mainzer}, {Ressler},
  {Cutri}, {Jarrett}, {Kirkpatrick}, {Padgett}, {McMillan}, {Skrutskie},
  {Stanford}, {Cohen}, {Walker}, {Mather}, {Leisawitz}, {Gautier}, {McLean},
  {Benford}, {Lonsdale}, {Blain}, {Mendez}, {Irace}, {Duval}, {Liu}, {Royer},
  {Heinrichsen}, {Howard}, {Shannon}, {Kendall}, {Walsh}, {Larsen}, {Cardon},
  {Schick}, {Schwalm}, {Abid}, {Fabinsky}, {Naes}, \& {Tsai}}]{WISE}
{Wright}, E.~L., {Eisenhardt}, P.~R.~M., {Mainzer}, A.~K., {et~al.} 2010, \aj,
  140, 1868

\end{thebibliography}

\end{document}